\RequirePackage{lineno}
\documentclass[aps,twocolumn,showpacs,byrevtex]{revtex4-1}
\usepackage{times}
\usepackage{amsmath}
\usepackage{graphicx}
\usepackage{color}
\usepackage[T1]{fontenc}
\usepackage[implicit=true]{hyperref}

\newcommand{\x}{\chi_{c1}(3872)}
\newcommand{\pp}{\pi^+\pi^-}

\newcommand{\LL}{\ell^+\ell^-}
\newcommand{\EE}{e^+e^-}

\newcommand{\GG}{\gamma\gamma}

\newcommand{\psip}{\psi(2S)}

\newcommand{\jpsi}{J/\psi}
\newcommand{\piz}{\pi^0}

\newcommand{\ppjpsi}{\pi^+\pi^-J/\psi}

\begin{document}

\title{\boldmath  Search for the light hadron decay $\chi_{c1}(3872) \to \pi^{+}\pi^{-}\eta$
}

\author{
\begin{small}
\begin{center}
M.~Ablikim$^{1}$, M.~N.~Achasov$^{5,b}$, P.~Adlarson$^{75}$, X.~C.~Ai$^{81}$, R.~Aliberti$^{36}$, A.~Amoroso$^{74A,74C}$, M.~R.~An$^{40}$, Q.~An$^{71,58}$, Y.~Bai$^{57}$, O.~Bakina$^{37}$
% ...
\\
\vspace{0.2cm}
(BESIII Collaboration)\\
\vspace{0.2cm} {\it
	$^{1}$ Institute of High Energy Physics, Beijing 100049, People's Republic of China\\
% ...
$^{5}$ Budker Institute of Nuclear Physics SB RAS (BINP), Novosibirsk 630090, Russia\\
% ...
$^{36}$ Johannes Gutenberg University of Mainz, Johann-Joachim-Becher-Weg 45, D-55099 Mainz, Germany\\
	$^{37}$ Joint Institute for Nuclear Research, 141980 Dubna, Moscow region, Russia\\
% ...
$^{40}$ Liaoning Normal University, Dalian 116029, People's Republic of China\\
% ...
$^{57}$ Southeast University, Nanjing 211100, People's Republic of China\\
	$^{58}$ State Key Laboratory of Particle Detection and Electronics, Beijing 100049, Hefei 230026, People's Republic of China\\
% ...
$^{71}$ University of Science and Technology of China, Hefei 230026, People's Republic of China\\
% ...
$^{74}$ University of Turin and INFN, (A)University of Turin, I-10125, Turin, Italy; (B)University of Eastern Piedmont, I-15121, Alessandria, Italy; (C)INFN, I-10125, Turin, Italy\\
	$^{75}$ Uppsala University, Box 516, SE-75120 Uppsala, Sweden\\
% ...
$^{81}$ Zhengzhou University, Zhengzhou 450001, People's Republic of China\\
	\vspace{0.2cm}
% ...
$^{b}$ Also at the Novosibirsk State University, Novosibirsk, 630090, Russia\\
% ...
}
\end{center}
\vspace{0.4cm}
\end{small}
}

\date{\today}

\begin{abstract}

  With a data sample corresponding to an integrated luminosity of 11.5~fb$^{-1}$ 
  collected with the BESIII detector operating at the BEPCII storage ring, for the first time the light hadron decay $\chi_{c1}(3872) \rightarrow  \pi^{+}\pi^{-}\eta$ 
  is searched for. While no significant signal is observed, the upper limits at the 90\% confidence level for 
  $\sigma[e^{+}e^{-} \rightarrow \gamma \chi_{c1}(3872)] \mathcal{B}[\chi_{c1}(3872) \rightarrow  \pi^{+}\pi^{-}\eta]$ at center-of-mass energies from 4.13 to 4.34 GeV are determined.
  By normalizing to the $\x\to\ppjpsi$ decay channel, a 90\% confidence level upper limit for the branching fraction ratio
  $\mathcal{R}=\mathcal{B}[\chi_{c1}(3872) \rightarrow\pi^{+}\pi^{-}\eta]/\mathcal{B}[\chi_{c1}(3872) \rightarrow \pi^{+}\pi^{-} J/\psi] < 0.12$ is given. 
  These measurements provide important inputs for understanding the internal structure of the $\x$ resonance.
\end{abstract}


\maketitle
\par Since its discovery in 2003~\cite{Belle:2003_firstfoundX}, the $\chi_{c1}(3872)$ resonance has been widely considered an exotic hadron state
beyond the conventional baryon and meson picture~\cite{RMP-Olsen}. At the moment, several theoretical models have been proposed that describe the $\chi_{c1}(3872)$
as a tetraquark state~\cite{Matheus:tetraquark}, a $P$-wave radially excited charmonium state~\cite{Barnes:chic1}, 
or a $D^{0}\bar{D}^{*0}$ meson molecule~\cite{Lee:DDmeson,RMP-molecule}, the latter one being favored due to the $\chi_{c1}(3872)$ mass being very close to the $D^{0}\bar{D}^{*0}$ mass threshold~\cite{ParticleDataGroup:2018ovx}.
However, the production rate of $\chi_{c1}(3872)$ in high energy $pp/p\bar{p}$ collisions is comparable to that of the $\psip$~\cite{pp-production}, which does not
agree with a pure molecule prediction~\cite{molecule-suppress}. To explain this, the $\x$ has been proposed to be a charmonium-hadronic molecule mixture state~\cite{mix}.

Many experimental measurements have investigated the
mass, width~\cite{ParticleDataGroup:2018ovx}, spin-parity~\cite{spinLHCb:2013kgk,spinLHCb:2015jfc} and 
decays~\cite{BESIII:DECAY_process1,BESIII:DECAY_process2,BESIII:DECAY_process3,BaBar:DECAY_process4,BESIII:DECAY_process5,Belle:DECAY_process6} of the $\x$. 
The branching fractions (BFs) of several final states involving charmonium or the $D^{(*)}$ meson,  
such as  $\pi^+\pi^-J/\psi,~\omega J/\psi,~\gamma J/\psi,~\pi^0\chi_{c1}(1P)$, and $D^0\bar{D}^{*0}$ 
have been  measured~\cite{ParticleDataGroup:2018ovx}. 
Yet, no decay final state without charm, such as light hadrons has been observed, except for an evidence of $\gamma \gamma^*$ production~\cite{Belle:DECAY_process7}. 
Theoretically, if the $\x$ was a $D^{0}\bar{D}^{*0}$ molecular state, its size would be very large ($\sim$ 10 fm)
due to a small binding energy~\cite{RMP-molecule} and the $c$ and $\bar{c}$ quarks would be expected to be far away from each other. 
Consequently, the light hadron decay of $\x$, which requires $c\bar{c}$ annihilation~\cite{Lee:DDmeson, swanson}, would be suppressed. 
If, however, the $\x$ wave function contains a sizeable $c\bar{c}$ core component, the BFs of these final states will be enhanced.

To further understand the nature of the $\chi_{c1}(3872)$, more experimental studies about its decay channels are required.
By taking measurements from several experiments into account, 
Ref.~\cite{Li:combined_X3872} has performed a global analysis on the BFs of its decays, which indicates that $(32^{+18}_{-32})$\% of the BFs have not yet been observed in experiment.
The BESIII experiment accumulated the world's largest $\EE$ annihilation data from $\sqrt{s}=4.01$ to 4.95 GeV and offers an opportunity to further investigate decays of $\chi_{c1}(3872)$, shedding light on the wave function of the $\chi_{c1}(3872)$ state.
Previously, the BESIII collaboration has already established the $e^{+}e^{-} \rightarrow \gamma \chi_{c1}(3872)$ production method with large significance~\cite{BESIII:DECAY_process2,BESIII:DECAY_process3}.

\par In this paper, based on the assumption that $\chi_{c1}(3872)$ contains a sizeable component of the $\chi_{\rm c1}(2P)$,
~we search for the light hadron decay process $\chi_{c1}(3872)\to\pp\eta$, where the $\x$ particle is produced via the radiative transition $\EE\to\gamma\x$.
The $\eta$ candidate is reconstructed via its $\GG$ and $\pp\piz(\to\GG)$ decays. A data sample taken at center-of-mass  $\rm (c.m.)$ energy from  $\sqrt{s}=4.13$ to 4.34 GeV~\cite{BESIII:data_2015,BESIII:data_2021} is used, 
corresponding to an integrated luminosity of 11.5 $\rm fb^{-1}$~\cite{BESIII:lum_2015,BESIII:lum_2017}, see Table~\ref{fitting results smear}.

The BESIII detector~\cite{Ablikim:BESIII_detector} records symmetric $e^+e^-$ collisions provided by the BEPCII storage ring~\cite{Yu:BEPCII}, which operates 
in the c.m. energy range from 2.0 to 4.95~GeV. BESIII has collected large data samples in this energy region~\cite{Ablikim:Physics_Programme}. The cylindrical core of the BESIII detector covers 93\% of the full solid angle and consists of a helium-based multilayer drift chamber~(MDC), a plastic scintillator time-of-flight system~(TOF), and a CsI(Tl) electromagnetic calorimeter~(EMC), which are all enclosed in a superconducting solenoidal magnet providing a 1.0~T  (0.9~T in 2012) magnetic field. The solenoid is supported by an octagonal flux-return yoke with resistive plate counter muon identification modules interleaved with steel. The charged-particle momentum resolution at $1~{\rm GeV}/c$ is $0.5\%$, and the ${\rm d}E/{\rm d}x$ resolution is $6\%$ for electrons from Bhabha scattering. The EMC measures photon energies with a resolution of $2.5\%$ ($5\%$) at $1$~GeV in the barrel (end cap) region. The time resolution in the TOF barrel region is 68~ps, while that in the end cap region is 110~ps.
The end-cap TOF
system was upgraded in 2015 using multi-gap resistive plate chamber
technology, providing a time resolution of
60~ps~\cite{etof}.

Simulated data samples produced with a {\sc geant4}-based~\cite{geant4} Monte Carlo (MC) package, which includes the geometric description of the BESIII detector and the
detector response, are used to determine detection efficiencies, optimize event selection, and estimate backgrounds. 
For the signal process, 100,000 $\EE\to\gamma\x$
events are generated at each c.m. energy, assuming an $E1$ radiative transition process which has been verified by BESIII data~\cite{BESIII:DECAY_process5}. The possible Initial-State-Radiation (ISR) is simulated with {\sc kkmc}~\cite{ref:kkmc},
 incorporating the $\sqrt{s}$-dependent production cross section of $\EE\to\gamma\x$ ~\cite{BESIII:DECAY_process5}. 
The maximum ISR photon energy is set according to the production threshold (3.9~GeV) of the $\gamma \chi_{c1}(3872)$ system.
%
The decay of $\chi_{c1}(3872) \rightarrow \pi^{+}\pi^{-}\eta$ and $\eta\to\GG$ is generated by {\sc evtgen}~\cite{ref:evtgen} with the phase-space model. The other decay $\eta \rightarrow \pi^{+}\pi^{-}\pi^{0}$ is generated according to the Dalitz-distribution measured by BESIII~\cite{BESIII:pi0_dalitz_distribution}.
The final-state-radiation of charged particles is simulated with the {\sc photos} package~\cite{photos}.

\par To study the possible background, an inclusive MC sample is used, including the production of open charm processes, the ISR production of vector charmonium(-like) states, and the continuum processes incorporated in {\sc kkmc}~\cite{ref:kkmc}. All particle decays are modeled with {\sc evtgen}~\cite{ref:evtgen} using BFs taken from the Particle Data Group (PDG)~\cite{ParticleDataGroup:2018ovx}, when available,
or otherwise modelled with {\sc lundcharm}~\cite{ref:lundcharm} for charmonium states and {\sc pythia}~\cite{pythia} for other hadrons. 
The equivalent integrated luminosity of the inclusive MC sample is 10 times that of data at $\sqrt{s} =4178.0$ MeV, and is of equal size to data at other c.m. energies.


\par Charged tracks detected in the MDC are required to be within a polar angle ($\theta$) range of $|\rm{cos\theta}|<0.93$, where $\theta$ is defined with respect to the $z$-axis, which is the symmetry axis of the MDC.  For each charged track, the distance of closest approach to the interaction point must be less than 10\,cm along the $z$-axis, $|V_{z}|$, and less than 1\,cm in the transverse plane, $|V_{xy}|$. Photon candidates are identified using showers in the EMC. The deposited energy of each shower must be more than 25~MeV in the barrel region ($|\cos\theta|< 0.80$) and more than 50~MeV in the end cap region ($0.86 <|\cos \theta|< 0.92$). To exclude showers that originate from charged tracks, the angle subtended by the EMC shower and the position of the closest charged track at the EMC must be greater than 10 degrees. To suppress electronic noise and showers unrelated to the event, the EMC time with respect to the event start time is required to be within [0, 700]~ns.

\par For the process of  $e^{+}e^{-} \rightarrow \gamma \chi_{c1}(3872)\rightarrow \gamma \pi^{+}\pi^{-}\eta$ with $\eta \to \gamma\gamma$, 
the number of charged tracks is required to be two with zero net charge in an event, and both tracks are assigned as pion candidates.
We also require at least three good photon candidates for each event.
To improve the resolution and reduce backgrounds, a five-constraint (5C) kinematic fit is performed. 
Four constraints come from the four-momentum conservation of the final state particles equal to the initial $\EE$ colliding beams, 
and the additional one comes from the $\eta$ mass. For possible multi-combination of photon candidates, 
the one with the minimum $\chi^{2}_{\rm 5C}$ is retained, and $\chi^{2}_{\rm 5C}<16$ is further required. This value is determined by optimizing the Figure-of-Merit (FOM) $S/\sqrt{S+B}$, where $S$ represents the number of signal events from signal MC simulation, and $B$ is
the number of background events estimated from the inclusive MC sample.

\begin{figure}[!htbp]
	\centering
	\includegraphics[width=0.45\textwidth, height=0.3\textwidth]{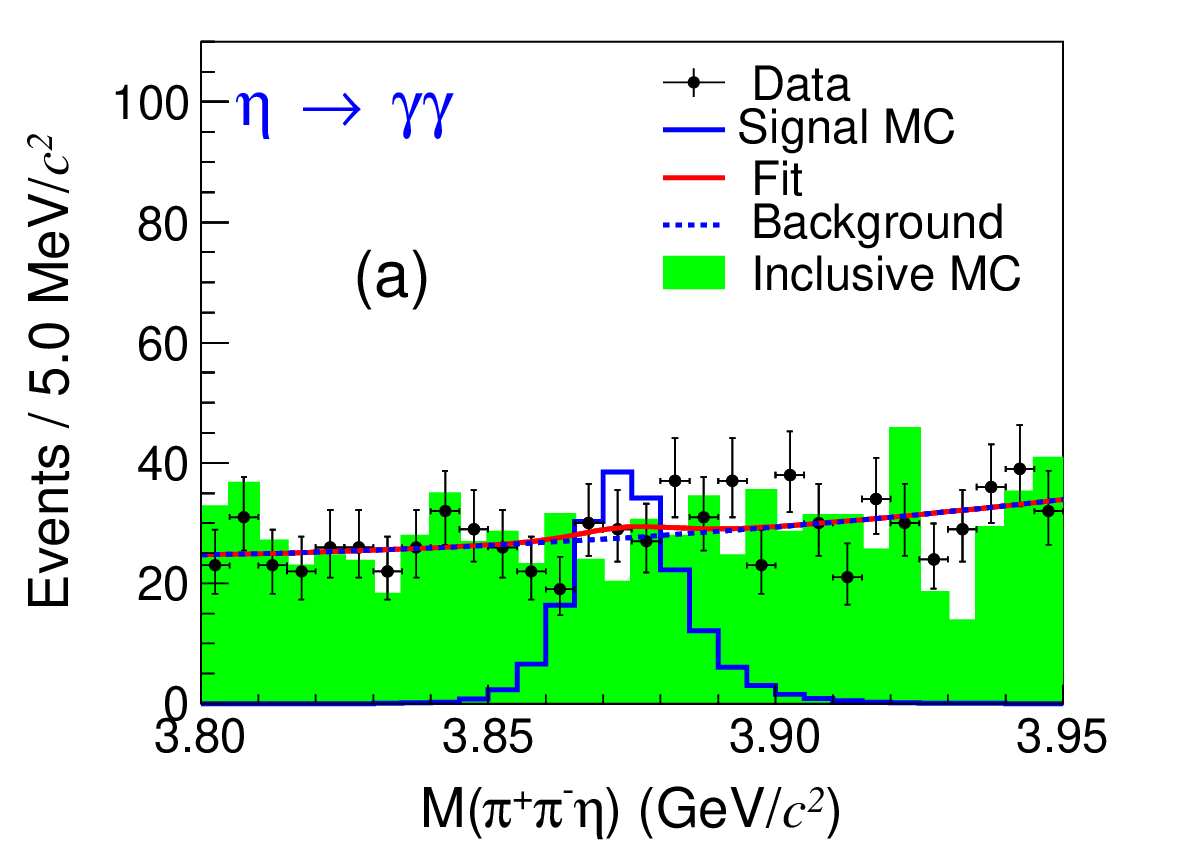}\\
	\includegraphics[width=0.45\textwidth, height=0.3\textwidth]{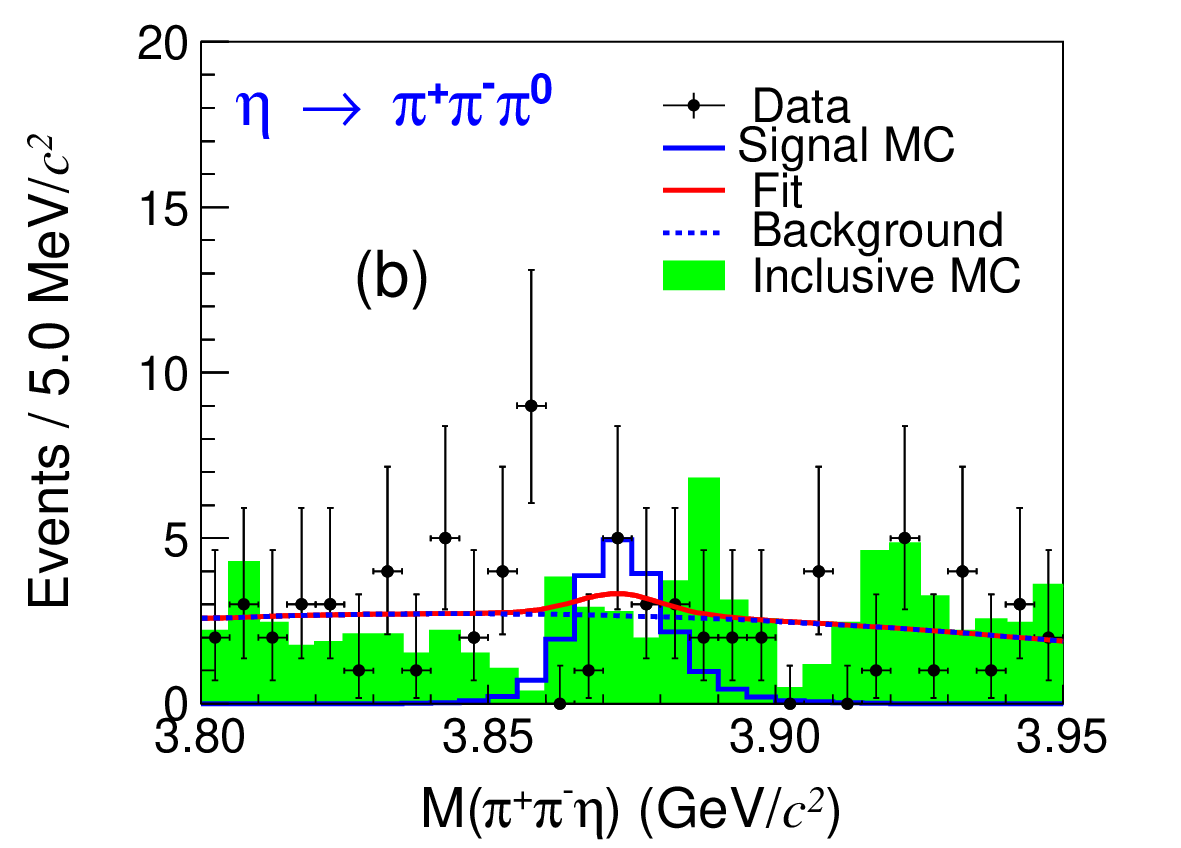}	
	\caption{The $M(\pp\eta)$ distributions from $\eta\to\GG$ mode (a) and $\pp\piz$ mode (b) with fit curves overlaid.
	Dots with error bars are data, the blue solid curves are the MC signal with arbitrary normalization, the red solid curves are the fit result, the blue dashed curves are the background components, 
	and the green shaded histograms are the backgrounds estimated from the inclusive MC sample.}
\label{comb_gamma_pipmpi0}
\end{figure}

By analysing the inclusive MC sample, we find that there are mis-combination backgrounds from $\pi^0\to\gamma\gamma$ decays.
To reduce these background events, $|M(\gamma_{\rm rad}\gamma_{1(2)})-m(\piz)|>20$~MeV/$c^{2}$ 
is required, where $\gamma_{\rm rad}$ refers to the radiative photon in $\EE\to\gamma\x$, $\gamma_{1(2)}$ denote the two photons from $\eta$ decay, $M(\gamma_{\rm rad}\gamma_{1(2)})$ is the invariant mass of $\gamma_{\rm rad}$ and $\gamma_{1(2)}$ combination, and $m(\piz)$ is the  nominal mass of $\piz$~\cite{ParticleDataGroup:2018ovx}. 
In addition, there are also possible background events from the radiative dimuon ($e^{+}e^{-} \rightarrow (\gamma) \mu^{+}\mu^{-}$) process, which is removed by requiring the opening angle between pion pair to fulfill $\cos\theta_{\pi^{+}\pi^{-}}>-0.96$. 
After imposing these event selection criteria, the distribution of $M(\pi^{+}\pi^{-}\eta)$ from the full data set is shown in Fig.~\ref{comb_gamma_pipmpi0}~(a).
No $\chi_{c1}(3872)$ signal is observed, and the background contribution estimated from the inclusive MC sample shows no peaking structure in the signal region.

\par  For the process of  $e^{+}e^{-} \rightarrow \gamma \chi_{c1}(3872)\rightarrow \gamma \pi^{+}\pi^{-}\eta$ with $\eta\to\pp\piz$ and $\piz \to \GG$, the number of charged tracks is required to be four with zero net charge in an event. Due to possible kaon background contamination, particle identification~(PID) for charged tracks, combining measurements of the ionization energy loss~(d$E$/d$x$) in the MDC
and the flight time in the TOF to evaluate the likelihoods $\mathcal{L}(h)~(h=p,K,\pi)$ for each hadron $(h)$ hypothesis, is applied.
At least one of the pion candidates is required to satisfy $\mathcal{L}(\pi)>\mathcal{L}(K)$.
We also require at least three good photon candidates for each event. 
To improve the resolution and reduce backgrounds, a 5C kinematic fit 
is performed, where
four constraints come from the four-momentum conservation of the final state particles equal to the initial $\EE$ colliding beams, 
and the additional one comes from the $\eta$ mass.
For possible multi-combination due to multi-pion and multi-photon candidates in an event, the one with the minimum $\chi^{2}_{\rm 5C}$ is retained
and $\chi^{2}_{\rm 5C}<30$ is further required. This value is determined by optimizing the FOM. 
The $\pi^{0}$ mass window is defined as $|M(\GG)-m(\piz)|<19.6$~MeV/$c^{2}$ ($\pm$3$\sigma$ around the $\piz$ mass).

Through the study of the inclusive MC sample, we find that the main backgrounds come from $e^{+}e^{-} \rightarrow\pi^{+}\pi^{-}\pi^{+}\pi^{-}\pi^{0}$.
Events satisfying $125.6<M(\gamma_{\rm rad}\gamma_{1})<150.0$ MeV/$c^{2}$ and $115.7<M(\gamma_{\rm rad}\gamma_{2})<160.0$ MeV/$c^{2}$ are rejected
to suppress this background, where $\gamma_{1(2)}$ denote the two photons from $\piz$ decay (sorted by energies)
and $M(\gamma_{\rm rad}\gamma_{1(2)})$ is the invariant mass of $\gamma_{\rm rad}$ and $\gamma_{1(2)}$ combination.
The different mass window vetos are optimized by FOM.
After imposing the selection criteria above, the obtained $M(\pi^{+}\pi^{-}\eta)$ distribution from the full data set is shown in Fig.~\ref{comb_gamma_pipmpi0} (b).
There is no obvious $\x$ signal observed, and the background level is lower than that of the $\eta\to\GG$ mode.
The background also produces a smooth distribution in the signal region from the study of the inclusive MC sample.



\par For the two decay channels of $\eta$, a simultaneous fit to the $M(\pp\eta)$ distribution is performed to extract the signal yield at each c.m. energy.
In the fit, the signal probability-density-function is parameterized with an MC simulated shape (with $\x$ mass
and width taken from PDG~\cite{ParticleDataGroup:2018ovx}), convolved with a Gaussian function which represents
the resolution difference between data and MC simulation. 
The mean and standard deviation of the Gaussian function are fixed to the values obtained from a control sample of $\chi_{c1}(1P)\to\pp\eta$. 
The background shape is parameterized as a 2nd-order polynomial function. 
In the simultaneous fit, the production rate for $\EE\to\gamma\x\to\gamma\pp\eta$ is a common parameter,  and the signal yields  for $\eta\to\GG$ and $\pp\piz$ 
modes are weighted according to the detection efficiencies and BFs. Appendix~\ref{fit_ecm} shows the fit results at each c.m. energy,
and the fit results to the full data set are shown in Fig.~\ref{comb_gamma_pipmpi0}. 

\par
\begin{figure}[!htbp]
	\centering
	\includegraphics[width=0.45\textwidth, height=0.3\textwidth]{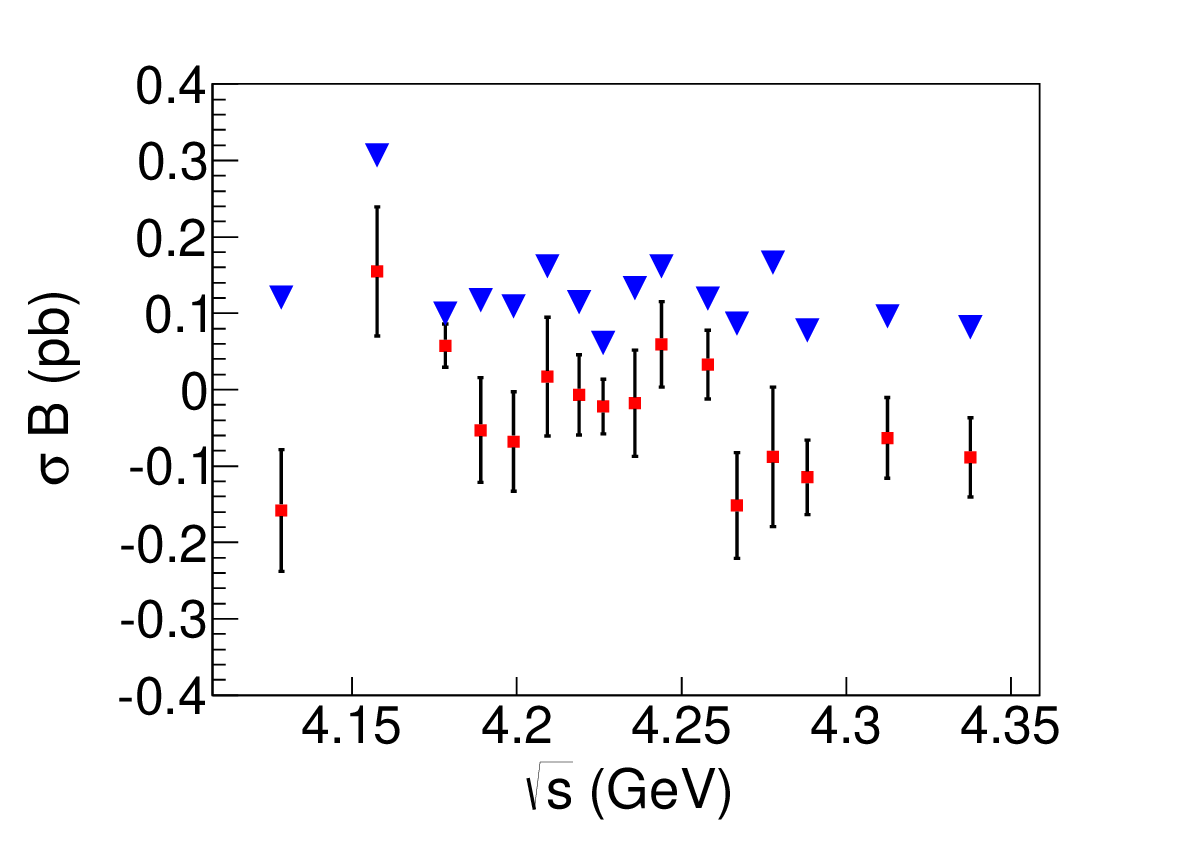}
	\caption{The production cross section of $e^{+}e^{-} \rightarrow \gamma \chi_{c1}(3872)$ times the BF of 
		$\chi_{c1}(3872) \rightarrow \pi^{+}\pi^{-}\eta$ as a function of center-of-mass energy. Blue triangles are the upper limit for $(\sigma  \mathcal{B})^{\rm up}$ at 90\% C.L., and the red dots with error bars are the nominal result.
	}
	\label{Cs_distribution_err}
\end{figure}

 The production cross section $\sigma[e^{+}e^{-}\rightarrow \gamma \chi_{c1}(3872)]$ times the BF
$\mathcal{B}[\chi_{c1}(3872)\rightarrow\pi^{+}\pi^{-}\eta]$ is calculated as
\begin{eqnarray}\notag
	\sigma[e^{+}e^{-}\rightarrow \gamma \chi_{c1}(3872)]\mathcal{B}[\chi_{c1}(3872)\rightarrow\pi^{+}\pi^{-}\eta] \\
	=\frac{N^{\rm obs}_i}{\mathcal{L}_{\rm int}  (1+\delta) ( \varepsilon_1  \mathcal{B}_1+ \varepsilon_2  \mathcal{B}_2)} 
	= \frac{N^{\rm sig}_i}{ \mathcal{L}_{\rm int}  (1+\delta)},  
\end{eqnarray}
where $N^{\rm obs}_i$ is the number of observed signal events from the $i$-th data set, 
$\mathcal{L}_{\rm int}$ is the corresponding integrated luminosity, $(1+\delta)$ is the radiative correction factor 
calculated by the {\sc kkmc}  MC generator~\cite{ref:kkmc}, $\varepsilon_j\mathcal{B}_j$~$(j=1,2)$ is the product of detection efficiency
and BF for the $\eta\to\GG$ and $\pp\piz$ modes, 
and $N^{\rm sig}_i$ is the number of produced signal events after considering the detection efficiencies and BFs.
Since no obvious signal is seen in the $M(\pi^{+}\pi^{-}\eta)$ distribution, an upper limit for $\sigma[\EE\to\gamma\x]\mathcal{B}[\x\to \pi^{+}\pi^{-}\eta]$ 
is set using the Bayesian approach~\cite{Bayesian:Data_Analysis}.
%
%
By scanning the likelihood function $\mathcal{L}(n)$ in the fit, the number of events corresponding to 90\% of the integral 
$\int_{0}^{N^{\rm up}}\mathcal{L}(n)dn=0.9$ ($\int_{0}^{+\infty}\mathcal{L}(n)dn=1$) is estimated as the upper limit ($N^{\rm up}$) 
for the signal yield. The systematic uncertainty is considered by applying a Gaussian (with standard deviation equal to the total systematic uncertainty) smearing to the likelihood function.
%
  %
The upper limits $(\sigma[e^{+}e^{-}\rightarrow\gamma \chi_{c1}(3872)]\mathcal{B}[\chi_{c1}(3872)\rightarrow\pi^{+}\pi^{-}\eta])^{\rm up}$ at 90\% confidence level (C.L.) are measured 
from $\sqrt{s}=$4128.8 to 4337.9~MeV and are shown in Figure~\ref{Cs_distribution_err} (corresponding
numerical results are listed in Table~\ref{fitting results smear}). 

\begin{table*}[ht]
	\caption{The production cross section of $\EE\to\gamma\x$ times $\mathcal{B}[\x\to\pp\eta]$.
	Here $\sqrt{s}$ is the c.m. energy of data set, $\mathcal{L}_{\rm int}$ is the integrated luminosity, $\varepsilon_{\eta\to\GG~(\pp\piz)}$ is detection efficiency for
	$\eta\to\GG~(\pp\piz)$ modes, $(1+\delta)$ is the radiative correction factor,
	$N_{\rm sig}$ is the efficiency and branching fraction corrected production yield
	of $\chi_{c1}(3872) \rightarrow \pi^{+}\pi^{-}\eta$ signal events in $e^{+}e^{-}\rightarrow\gamma\chi_{c1}(3872)$ and $N_{\rm sig}^{\rm up}$ is the corresponding upper limit at the 90\% C.L., 
	$(\sigma \mathcal{B})$ is the product of the cross section $\sigma[e^{+}e^{-} \rightarrow \gamma \chi_{c1}(3872)]$ 
	and BF $\mathcal{B}[\chi_{c1}(3872) \rightarrow \pi^{+}\pi^{-}\eta]$ and
	$(\sigma \mathcal{B})^{\rm up}$ is the corresponding upper limit at the 90\% C.L.,
	and $\Delta$ denotes the total systematic uncertainty in the cross section measurement.}
	\centering
	\begin{tabular}{ c c c c c c c c c c}
		\hline
		\hline
		 $\sqrt{s}$~(MeV) & $\mathcal{L}_{\rm int}~(\rm pb^{-1})$ & $\varepsilon_{\eta\rightarrow\gamma\gamma} $ (\%)& $\varepsilon_{\eta\rightarrow\pi^{+}\pi^{-}\pi^{0}} $ (\%)& $(1+\delta)$& $N_{\rm sig}$ & $(\sigma  \mathcal{B})~(\rm pb)$ & $N_{\rm sig}^{\rm up}$ &  $(\sigma  \mathcal{B})^{\rm up}~(\rm pb)$  & $\Delta$ (\%)\\
		\hline
		 4128.8$\pm$0.4 & 401.5  & 23.1 & 13.3 & 0.78& -49.68 $\pm$ 25.07 & -0.16 $\pm$ 0.08 &38.0 & 0.12 & 7.2\\
		 4157.8$\pm$0.3 & 408.7  & 23.7 & 13.5 & 0.78& 49.08  $\pm$ 26.87 & 0.15  $\pm$ 0.08 &97.2 & 0.31 & 7.2\\
		 4178.0$\pm$0.8 & 3194.5 & 22.7 & 13.2 & 0.78& 143.08 $\pm$ 70.28 & 0.06  $\pm$ 0.03 &247.0& 0.10 & 7.6\\
		 4189.1$\pm$0.3 & 570.0  & 23.0 & 13.5 & 0.79& -23.70 $\pm$ 30.72 & -0.05 $\pm$ 0.07 &52.5 & 0.12 & 7.0\\
		 4199.2$\pm$0.3 & 526.7  & 23.3 & 13.8 & 0.80& -28.59 $\pm$ 27.34 & -0.07 $\pm$ 0.06 &46.0 & 0.11 & 7.8\\
		 4209.4$\pm$0.3 & 572.1  & 22.8 & 13.4 & 0.82&  7.94  $\pm$ 36.40 & 0.02  $\pm$ 0.08 &75.6 & 0.16 & 6.9\\
		 4218.9$\pm$0.3 & 569.2  & 22.7 & 13.5 & 0.84& -3.12  $\pm$ 25.11 & -0.01 $\pm$ 0.05 &54.9 & 0.12 & 7.9\\
		 4226.3$\pm$0.7 & 1100.9 & 22.6 & 13.4 & 0.86& -20.92 $\pm$ 34.04 & -0.02 $\pm$ 0.04 &58.2 & 0.06 & 7.5\\
		 4235.8$\pm$0.3 & 530.3  & 21.7 & 13.2 & 0.89& -8.36  $\pm$ 32.68 & -0.02 $\pm$ 0.07 &62.7 & 0.13 & 7.8\\
		 4244.0$\pm$0.3 & 594.0  & 21.4 & 12.8 & 0.92& 32.47  $\pm$ 30.57 & 0.06  $\pm$ 0.06 &88.2 & 0.16 & 7.6\\
		 4258.0$\pm$0.7 & 828.4  & 20.3 & 12.4 & 0.97& 26.36  $\pm$ 35.95 & 0.03  $\pm$ 0.04 &96.0 & 0.12 & 6.9\\
		 4266.8$\pm$0.3 & 529.7  & 19.9 & 11.9 & 1.00& -80.55 $\pm$ 36.76 & -0.15 $\pm$ 0.07 &46.0 & 0.09 & 6.7\\
		 4277.8$\pm$0.5 & 175.7  & 18.8 & 11.4 & 1.05& -16.10 $\pm$ 16.78 & -0.09 $\pm$ 0.09 &30.6 & 0.17 & 6.2\\
		 4288.4$\pm$0.3 & 498.5  & 18.8 & 11.0 & 1.09& -62.17 $\pm$ 26.29 & -0.11 $\pm$ 0.05 &42.0 & 0.08 & 7.7\\
		 4312.7$\pm$0.4 & 499.2  & 17.2 & 10.2 & 1.18& -37.23 $\pm$ 31.11 & -0.06 $\pm$ 0.05 &56.7 & 0.10 & 7.4\\
		 4337.9$\pm$0.4 & 511.5  & 16.1 & 9.3  & 1.28& -58.07 $\pm$ 33.89 & -0.09 $\pm$ 0.05 &53.7 & 0.08 & 7.2\\	
		\hline
        \hline		 
	\end{tabular}
	\label{fitting results smear}
\end{table*}


\par The systematic uncertainties in the cross section measurement come from integrated luminosity measurements, photon detection, tracking efficiencies, PID, $\eta/\pi^{0}$ reconstruction, kinematic fit, quoted BFs, radiative correction factor, signal extraction, and MC decay model.

%
The integrated luminosities of the data sets are measured using large-angle Bhabha scattering events, with a systematic uncertainty of 0.66\%~\cite{Yifan_Yang:Syslum}.
The photon detection uncertainty is estimated to be 1.0\% per photon by studying $\jpsi\to\rho\pi$ events~\cite{BESIII:Sysphoton1}.
%
The tracking efficiency uncertainty is estimated to be 1.0\% per pion by studying $\jpsi\to p\bar{p}\pp$ events~\cite{track}.
%
For the decay with $\eta\rightarrow\pi^{+}\pi^{-}\pi^{0}$, we use PID for the pion selection,
and at least one of the pion candidates is required to be identified. 
The PID efficiency of pions can be calculated as $1-(1-p)^{4}$, where $p$ represents the PID efficiency for a single pion.
Since $p$ is very high ($>0.8$)  at BESIII, the uncertainty due to PID can be safely ignored~\cite{BESIII:PID_pion_eff}.

%
The systematic uncertainty due to $\eta/\pi^{0}$ reconstruction is estimated to be 1.0\% per $\eta/\piz$, by studying a high purity control sample of $J/\psi \rightarrow \overline{p} p\eta$ ($\overline{p} p \pi^{0}$) events ~\cite{BESIII:Sys_reconstruction}. 
%
For the uncertainty from the kinematic fit, we use a helix parameter correction method~\cite{kinematic}. 
A correction is performed on the track parameters 
and half of the efficiency difference with and without the correction is taken as the systematic uncertainty. 

%
For the radiative correction factor calculation, the cross section of $\EE\to\gamma\x$ is input from Ref.~\cite{BESIII:DECAY_process5},
where a Breit-Wigner resonance $Y(4200)$ ($M[Y(4200)]=4200.6^{+7.9}_{-13.3}\pm3.0$ MeV/$c^2$ and $\Gamma[Y(4200)]=115^{+38}_{-26}\pm 12$ MeV) is employed to describe the cross section line shape.
To estimate the potential effect due to the uncertainty of the resonance, a two-dimensional Gaussian sampling (possible correlation has been considered)
method is used. We generate 300 groups of $Y(4200)$ resonance parameters (mass and width) and recalculate the $(1+\delta)\varepsilon$.
With the weighting method given in Ref.~\cite{Sun:ISR_iterative}, we get the distribution of $(1+\delta)  \varepsilon$ according to the Gaussian sampling.
A fit to this distribution yields the standard deviation of $(1+\delta) \varepsilon$, 
which is taken as the corresponding systematic uncertainty. 

%
In the nominal analysis, we use pure phase space to model the $\chi_{c1}(3872) \rightarrow \pi^{+}\pi^{-}\eta$ decay.
As $\x$ is a $J^{PC}=1^{++}$ resonance, we also try to model its decay by referring to a $1^{++}$ charmonium state $\chi_{c1}(1P)$. By selecting a pure control sample
$\psip\to\gamma\chi_{c1}(1P)$ with $\chi_{c1}(1P) \rightarrow \pi^{+}\pi^{-}\eta$ from the BESIII data~\cite{BESIII:control_chic1}, we take the decay model of $\chi_{c1}(1P)\to\pp\eta$ as 
an alternative, and the efficiency difference compared to the phase space model is taken as the associated systematic uncertainty. 
The BFs of $\eta/\piz$ decays are taken from PDG~\cite{ParticleDataGroup:2018ovx}, and the uncertainties are 0.5\% for $ \mathcal{B} (\eta \rightarrow \gamma\gamma)$, 1.22\% for $\mathcal{B} (\eta \rightarrow \pi^{+}\pi^{-}\pi^{0})$, and 0.03\% for $ \mathcal{B} (\pi^{0}\rightarrow\gamma\gamma)$, respectively.

%
The systematic uncertainty due to signal extraction is dominated by the signal and background shapes. 
For the signal shape, the $\x$ is simulated with resonance parameters (mass and width) taken from PDG. We vary the
resonance parameters by $\pm 1\sigma$ in the simulation. The background shape is varied
from a first-order polynomial to a second-order polynomial. By repeating these fits, we conservatively take the largest upper limit for signal yield as the final result. 
%
\par Appendix~\ref{sys_ecm} summarizes the systematic uncertainties at each c.m. energy. To combine the systematic uncertainties between
two $\eta$ decay modes, a weighted average is taken as
\begin{linenomath*}
\begin{equation}
	\Delta^2_\mathrm{tot}=\sum_{i=1}^{2}\omega_i^2\Delta_i^2+2\sum_{i\neq j}^{2} \rm cov(\it{i},\it{j}),
\end{equation}

\begin{equation}
	\omega_i = \frac{\epsilon_i\mathcal{B}_i}{\sum_{i=1}^2\epsilon_i\mathcal{B}_i},~\rm cov(\it{i},\it{j})=\rho_{\it{i}\it{j}}\omega_i\omega_j\Delta_i\Delta_j,
\end{equation}
\end{linenomath*}
where $\Delta_{\rm tot}$ is the combined total systematic uncertainty for each source, 
$\omega_i$ and $\Delta_i$ are the corresponding weight and systematic uncertainty, 
$\epsilon_i$ and $\mathcal{B}_i$ are the efficiency and
BF for the $i$-th decay mode of $\eta$, 
and $\rho_{ij}$ is the correlation coefficient between them. 
We take $\rho_{ij}=1$ for the same systematic source,
otherwise $\rho_{ij}=0$.

Assuming all these sources are independent, the total systematic uncertainties are added in quadrature for different sources,
which is around 7\%. 
Table~\ref{fitting results smear} lists the total systematic uncertainties in the cross section measurement.


\par As the $\EE\to\gamma\x\to\gamma\ppjpsi$ process has been observed 
at BESIII~\cite{BESIII:DECAY_process2,BESIII:DECAY_process3,BESIII:DECAY_process5}, 
we also measure the BF ratio 
$\mathcal{R}=\frac{\mathcal{B}[\x\to\pp\eta]}{\mathcal{B}[\x\to\ppjpsi]}$, which is calculated as
\begin{linenomath*}
\begin{equation}
\mathcal{R}=\frac{N_{\rm total}}{N_{\rm total}^{'}}\frac{\sum_{i}\mathcal{L}_{i}\sigma_i(1+\delta_{i})\epsilon_{i}^{'}\mathcal{B}(\jpsi\to\LL)}{\sum_{i}\mathcal{L}_{i}\sigma_i(1+\delta_{i})(\epsilon_{i}^{\GG}\mathcal{B}_1+\epsilon_{i}^{\pp\piz}\mathcal{B}_2)}
\label{br-ratio}
\end{equation}
\end{linenomath*}
where $N_{\rm total}$ is the total number of observed $\x$ signal events,
$\sigma_{i}$ is the $\EE\to\gamma\x$ production cross section~\cite{BESIII:DECAY_process5}, 
$\mathcal{L}_{i}$, $(1+\delta_{i})$, and $\epsilon_{i}$ are the integrated luminosity, radiative correction factor, and efficiency at the $i$-th c.m. energy, 
respectively~(cf. Table~\ref{fitting results smear}),
and $\mathcal{B}_{1,2}$ are BFs of $\eta \rightarrow \gamma\gamma$ and $\eta \rightarrow \pp\piz$.  The unprimed variables are for 
$\x\to\pp\eta$ and the primed variables are for $\x\to\ppjpsi$.  
Note that the integrated luminosity, the $\EE\to\gamma\x$ production cross section $\sigma_i$, 
and the radiative correction factor are the same for both processes.

\begin{figure}[!htbp]
\centering
   \includegraphics[width=0.45\textwidth, height=0.3\textwidth]{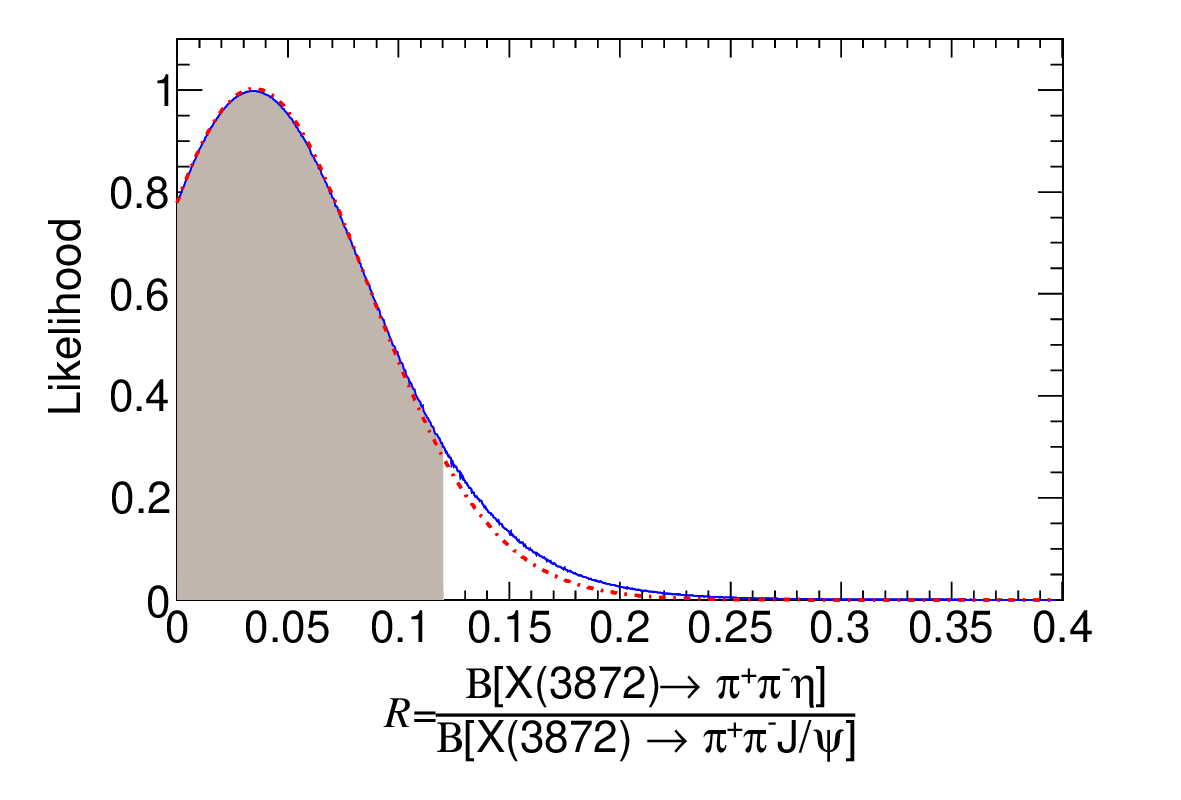}\hspace{5pt}
	\caption{The likelihood curve obtained by scanning various $\mathcal{R}$ values in the fit.
	The red dot-dashed curve is from the nominal fit, the blue solid curve is the corresponding distribution smeared by a Gaussian function with resolution equal to the total systematic uncertainty, and the gray shaded area indicates 90\% of the likelihood integral.
	}
	\label{UP_total}
\end{figure}

Since we do not observe any obvious signal in the $\x\to\pp\eta$ decay,
an upper limit at 90\% C.L. is determined for the BF ratio $\mathcal{R}$. 
For the normalization channel $\x\to\ppjpsi$,
the total number of signal events ($N_{\rm total}^{'}=86\pm10$) and the corresponding efficiencies are taken from Ref.~\cite{BESIII:DECAY_process5}.
The upper limit on $\mathcal{R}$ at the 90\% C.L. is defined as $\mathcal{R}^{\rm up}$, corresponding to the value at 90\% of the integral of the likelihood curve. 
The systematic uncertainty (cf. Table~\ref{Sum of sys err branching fraction}) is also taken into account by applying a Gaussian smearing to the likelihood curve.
The measured BF ratio is $\mathcal{R}=0.04 \pm 0.06$, and the corresponding
upper limit is set to $\mathcal{R}<0.12$, as shown in Fig.~\ref{UP_total}.

\par In the BF ratio measurement, systematic uncertainties including the integrated luminosity, $\EE\to\gamma\x$ production cross section, radiative correction factor, tracking efficiency and photon detection, are cancelled. The remaining systematic uncertainties come from extra photon and track detection, BF, $\eta$ ($\pi^{0}$) reconstruction, $J/\psi$ mass window, MC decay model, signal extraction, and the uncertainty of $\x$ signal events in the $\pi^{+}\pi^{-}J/\psi$ mode.

%
\par In the $\chi_{c1}(3872) \rightarrow \pi^{+}\pi^{-}\eta$ decay, two more photons are detected. A 1.0\% uncertainty is assigned to each photon~\cite{BESIII:Sysphoton1}, and the systematic uncertainty for photon detection is 2.0\%.
%
Similarly, there are only two charged tracks in the $\x\to\pp\eta(\to\GG)$ decay, while there are four in the $\x\to\ppjpsi(\to\LL)$ decay.
A 1.0\% uncertainty~\cite{track} is assigned for each track, and in total 2.0\% for the $\eta\to\GG$ mode.
%

%
The uncertainties from $\mathcal{B}(\eta\rightarrow\gamma\gamma)$, $\mathcal{B}(\eta\rightarrow\pi^{+}\pi^{-}\pi^{0})$, $\mathcal{B}(\pi^{0}\rightarrow\gamma\gamma)$,
and $\mathcal{B}(J/\psi\rightarrow l^{+}l^{-})$ are not cancelled out in the $\mathcal{R}$ measurement. 
Systematic uncertainties are assigned according to PDG~\cite{ParticleDataGroup:2018ovx}. 
%
The systematic uncertainty due to $\eta$ ($\pi^{0}$) reconstruction is also not cancelled out, and its contribution is 1.0\% as mentioned before.
%
The uncertainties due to the $\x\to\pp\eta$ decay model and $\x$ signal extraction are also discussed in the cross section measurement.
%
In the $\x\to\ppjpsi$ decay, the systematic uncertainty due to the $J/\psi$ mass window is 1.6\%~\cite{BESIII:DECAY_process5}.
%
Other uncertainties for the $\x\to\pi^{+}\pi^{-}J/\psi$ decay include the fit model, the background shape, and the statistical uncertainty of the signal yield, which have been studied in Ref.~\cite{BESIII:DECAY_process5} and the combined contribution is 12.3\%.

\par All the systematic uncertainties for the $\mathcal{R}$ measurement are listed in Table~\ref{Sum of sys err branching fraction}.
The contributions from $\eta\to\GG$ and $\pp\piz$ modes are combined using the same method as the cross section measurement.
Assuming all these sources are independent, the total systematic uncertainty is calculated by adding each individual source in quadrature.

\begin{table}[!htbp]
	\caption{Systematic uncertainties for the relative branching fraction $\mathcal{R}$ measurement.}
	\hspace{15pt}
	\centering
	\begin{tabular}{c c c c c c }
		\hline
		\hline
		Source & $\gamma\gamma$ &  &   & $ \pi^{+}\pi^{-}\pi^{0}$ & $\Delta$ (\%)\\
		\hline
        Photon   &\multicolumn{4}{c}{2.0} &2.0\\		
		Tracking   & 2.0 &  &   & 0.0 & 1.5\\
		$\mathcal{B}(\eta)$ &0.5&  &   &1.2&0.7\\
		$\mathcal{B}(J/\psi\rightarrow l^{+}l^{-})$ & \multicolumn{4}{c}{0.6} & 0.6\\
		$J/\psi$ mass window & \multicolumn{4}{c}{1.6} &  1.6\\  
		Decay model  & 7.7 &  &   & 3.9 & 6.7 \\
		$\eta$ ($\pi^{0}$) reconstruction &\multicolumn{4}{c}{1.0} &1.0\\
		$\pi^+\pi^{-}J/\psi$ yield & \multicolumn{4}{c}{12.3} & 12.3\\
		\hline
		Total &  &  &  &   &14.4\\
		\hline
		\hline
	\end{tabular}
	\label{Sum of sys err branching fraction}
\end{table}


\par In summary, we have searched for the decay $\chi_{c1}(3872) \rightarrow \pi^{+}\pi^{-}\eta$ associated with the radiative production 
$e^{+}e^{-} \rightarrow \gamma \chi_{c1}(3872)$ with 11.5~fb$^{-1}$ data at BESIII, at c.m. energies ranging from $\sqrt{s}$ = 4128.8 to 4337.9~MeV.
No significant signal has been observed and the upper limits for the product of the production cross section and BF $\sigma[e^{+}e^{-} \rightarrow \gamma \chi_{c1}(3872)]  \mathcal{B}[\chi_{c1}(3872) \rightarrow \pi^{+}\pi^{-}\eta]$ at 90\% C.L. have been given. We have also measured the BF ratio 
$\mathcal{R}=\frac{\mathcal{B}[\chi_{c1}(3872)\rightarrow\pi^{+}\pi^{-}\eta]}{\mathcal{B}[\chi_{c1}(3872)\rightarrow\pi^{+}\pi^{-}J/\psi]}$, and set an upper limit $\mathcal{R}\textless~0.12$ at 90\% C.L..
Taking $\mathcal{B}[\x\to\ppjpsi]=(3.8\pm1.2)\%$~\cite{ParticleDataGroup:2018ovx} as input, we obtain $\mathcal{B}[\x\to\pp\eta]<0.6\%$ at 90\% C.L..
A recent calculation assuming the $\x$ is a $DD^*$ molecule state shows that the light hadron decay width of $\x$ is in the range of $10$ to $10^2$~keV ~\cite{Theory_X_charmless}
[corresponding to BF $\sim$(1$-$10)\%]. Since our measurement obviously finds the BF to be lower than 1\%, this could have an impact
on the explanation of the internal structure of the $\chi_{c1}(3872)$ .


The BESIII Collaboration thanks the staff of BEPCII and the IHEP computing center for their strong support. This work is supported in part by National Key R\&D Program of China under Contracts Nos. 2020YFA0406300, 2020YFA0406400; National Natural Science Foundation of China (NSFC) under Contracts Nos. 11975141, 11635010, 11735014, 11835012, 11935015, 11935016, 11935018, 11961141012, 12022510, 12025502, 12035009, 12035013, 12061131003, 12192260, 12192261, 12192262, 12192263, 12192264, 12192265, 12221005, 12225509, 12235017; the Chinese Academy of Sciences (CAS) Large-Scale Scientific Facility Program; the CAS Center for Excellence in Particle Physics (CCEPP); Joint Large-Scale Scientific Facility Funds of the NSFC and CAS under Contract No. U1832207; CAS Key Research Program of Frontier Sciences under Contracts Nos. QYZDJ-SSW-SLH003, QYZDJ-SSW-SLH040; 100 Talents Program of CAS; Project ZR2022JQ02 supported by Shandong Provincial Natural Science Foundation; 
Hubei University of Automotive Technology under Contract No. BK202318;
The Institute of Nuclear and Particle Physics (INPAC) and Shanghai Key Laboratory for Particle Physics and Cosmology; ERC under Contract No. 758462; European Union's Horizon 2020 research and innovation programme under Marie Sklodowska-Curie grant agreement under Contract No. 894790; German Research Foundation DFG under Contracts Nos. 443159800, 455635585, Collaborative Research Center CRC 1044, FOR5327, GRK 2149; Istituto Nazionale di Fisica Nucleare, Italy; Ministry of Development of Turkey under Contract No. DPT2006K-120470; National Research Foundation of Korea under Contract No. NRF-2022R1A2C1092335; National Science and Technology fund of Mongolia; National Science Research and Innovation Fund (NSRF) via the Program Management Unit for Human Resources \& Institutional Development, Research and Innovation of Thailand under Contract No. B16F640076; Polish National Science Centre under Contract No. 2019/35/O/ST2/02907; The Swedish Research Council; U. S. Department of Energy under Contract No. DE-FG02-05ER41374.


%
%
\clearpage
\begin{widetext}
\appendix
\section{Fit to the $M(\pp\eta)$ mass distribution at each c.m. energy}
\label{fit_ecm}
\begin{figure*}[ht]
	\centering
\includegraphics[width=0.48\textwidth, height=0.21\textwidth]{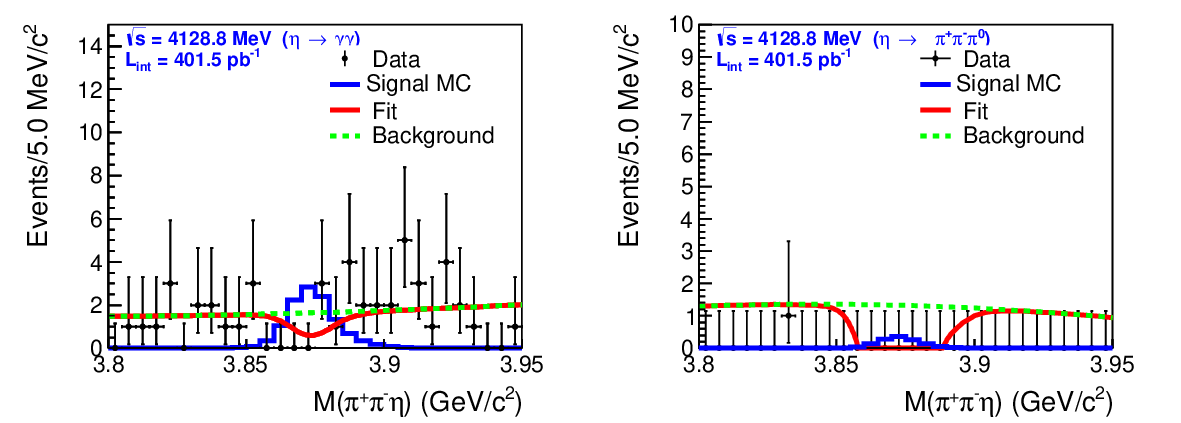}
\includegraphics[width=0.48\textwidth, height=0.21\textwidth]{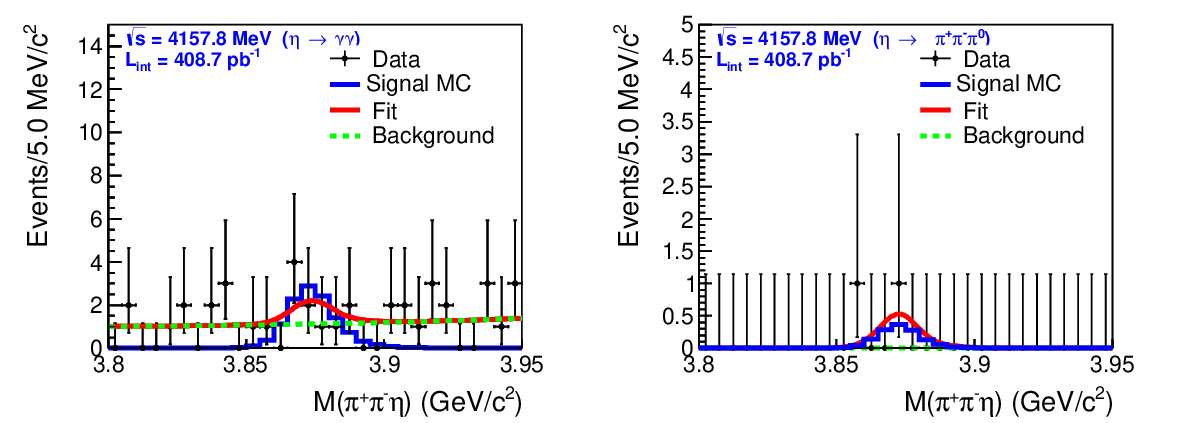}
\includegraphics[width=0.48\textwidth, height=0.21\textwidth]{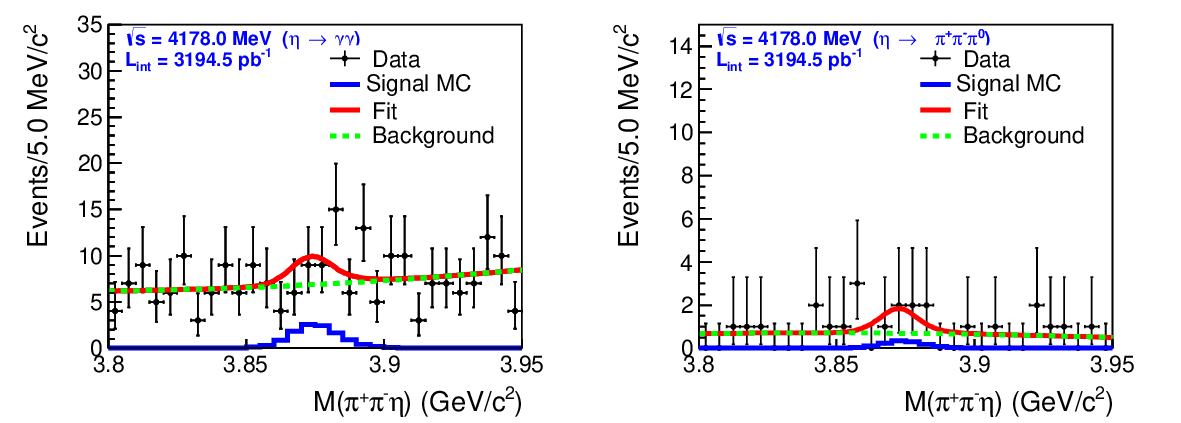}
\includegraphics[width=0.48\textwidth, height=0.21\textwidth]{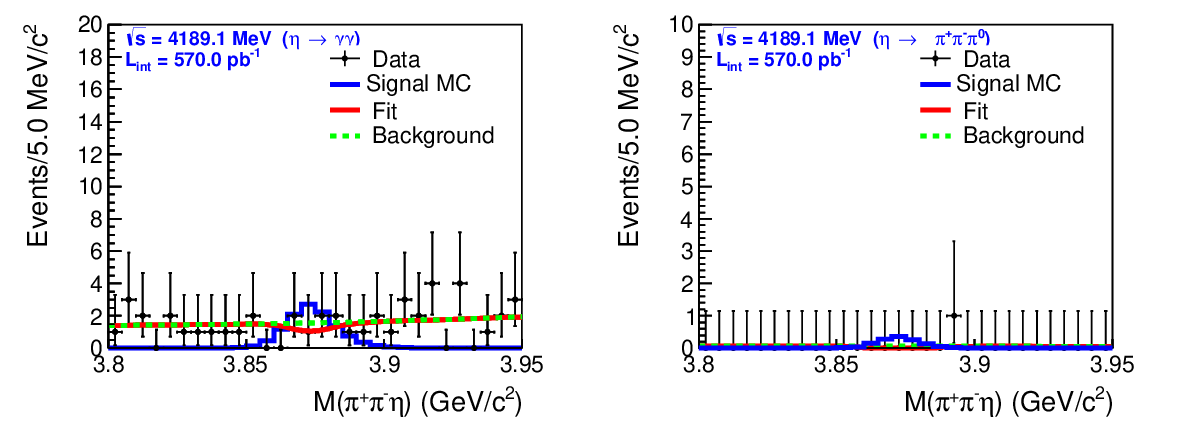}
\includegraphics[width=0.48\textwidth, height=0.21\textwidth]{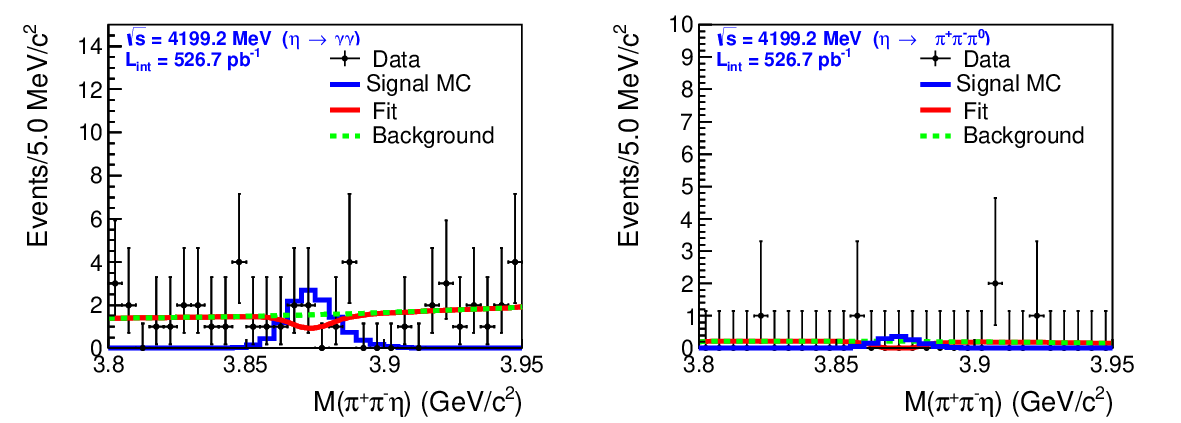}
\includegraphics[width=0.48\textwidth, height=0.21\textwidth]{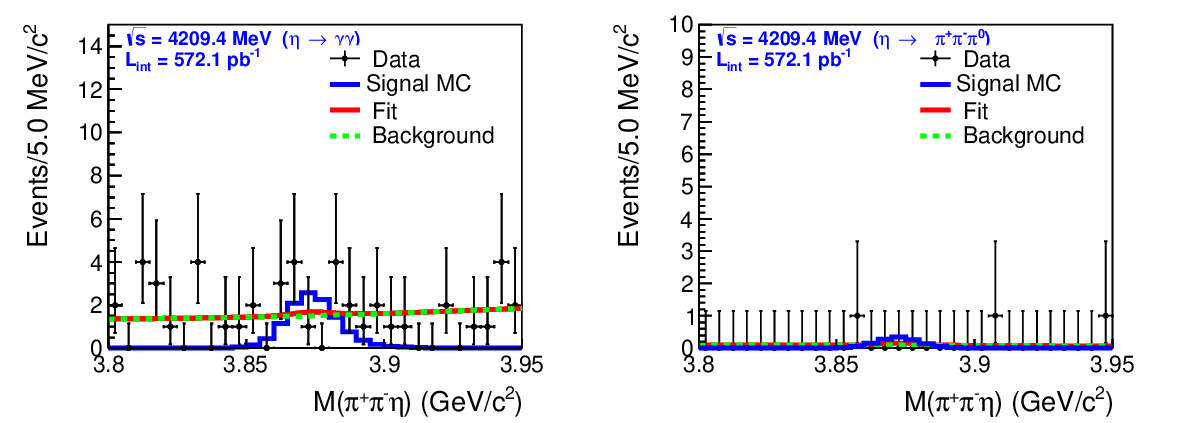}
\includegraphics[width=0.48\textwidth, height=0.21\textwidth]{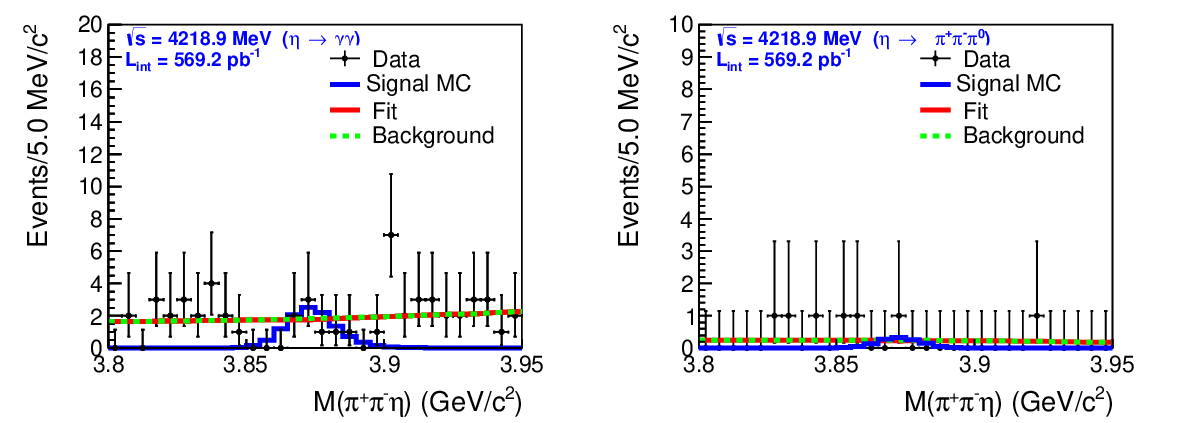}
\includegraphics[width=0.48\textwidth, height=0.21\textwidth]{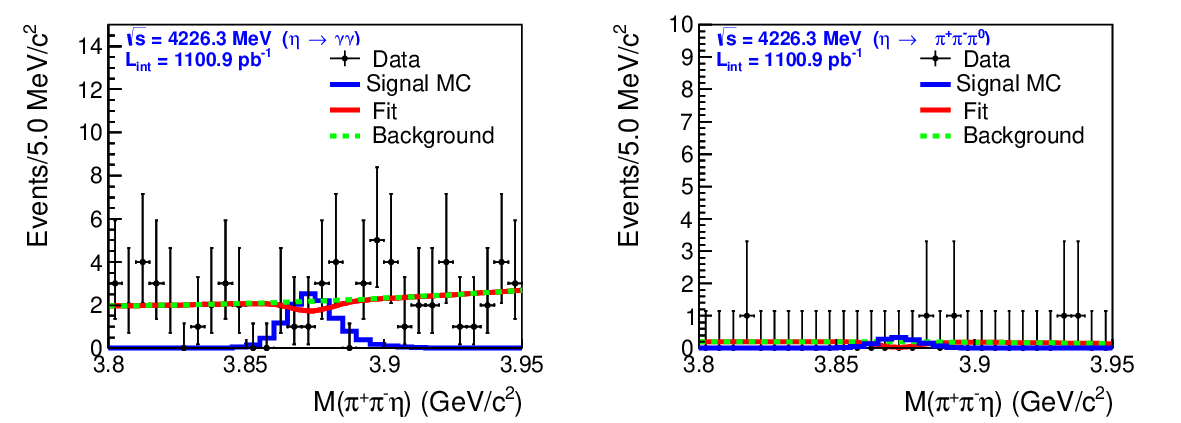}
\includegraphics[width=0.48\textwidth, height=0.21\textwidth]{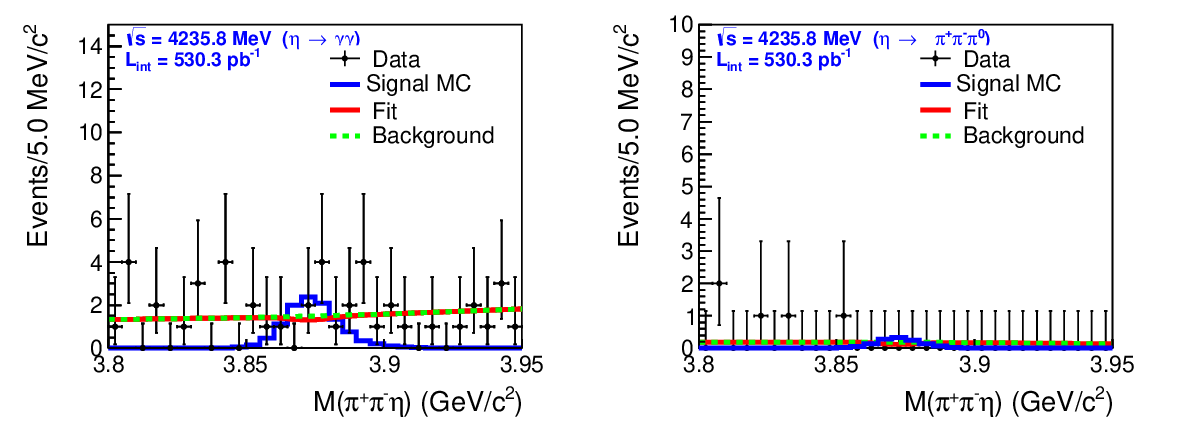}
\includegraphics[width=0.48\textwidth, height=0.21\textwidth]{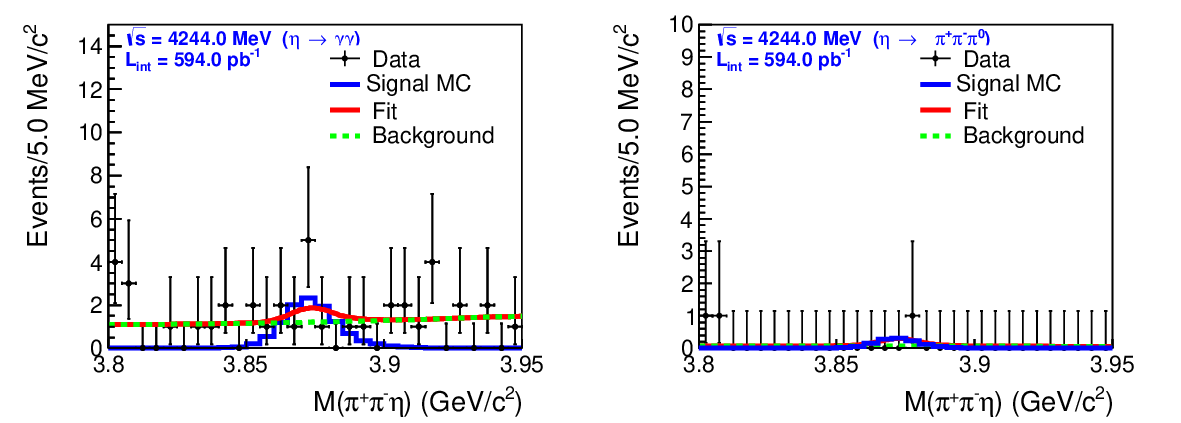}			
\caption{The fit results at c.m. energies ranging from $\sqrt{s}=$ 4128.8 to 4244.0 MeV. The signal MC shape convolved with a Gaussian function is used to fit the invariant mass spectrum of $\pi^{+}\pi^{-}\eta$ at each energy. Black dots with error bars are data, the red solid curve shows the fit result, the blue dashed curve is the signal shape with arbitrary normalization and the green dashed curve is the background contribution.}
\label{fit fig1}
\end{figure*}	
\begin{figure*}[ht]
	\centering			
	\includegraphics[width=0.48\textwidth, height=0.21\textwidth]{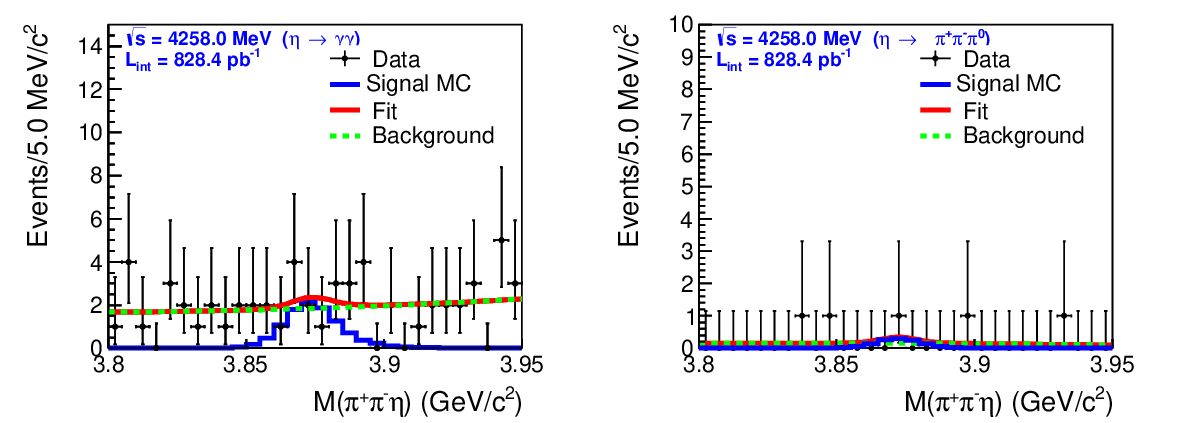}
	\includegraphics[width=0.48\textwidth, height=0.21\textwidth]{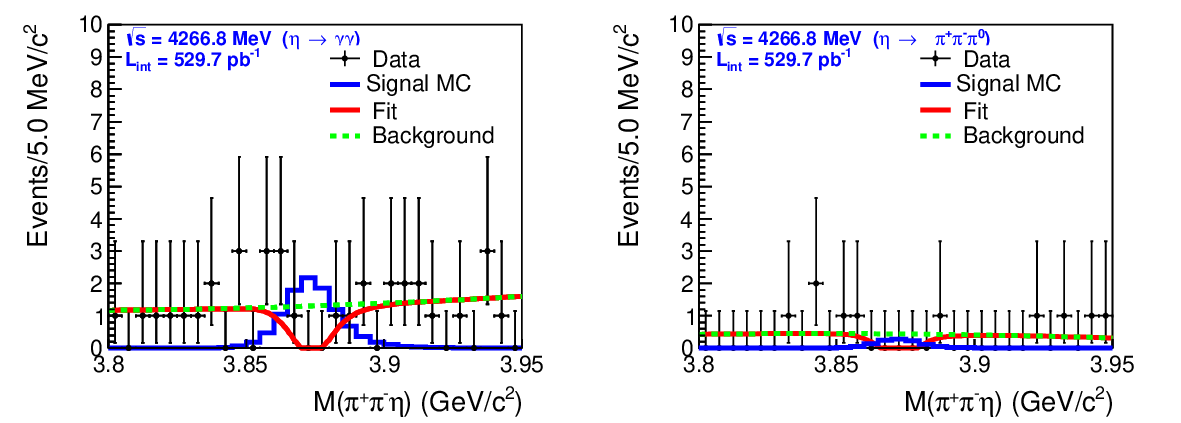}
	\includegraphics[width=0.48\textwidth, height=0.21\textwidth]{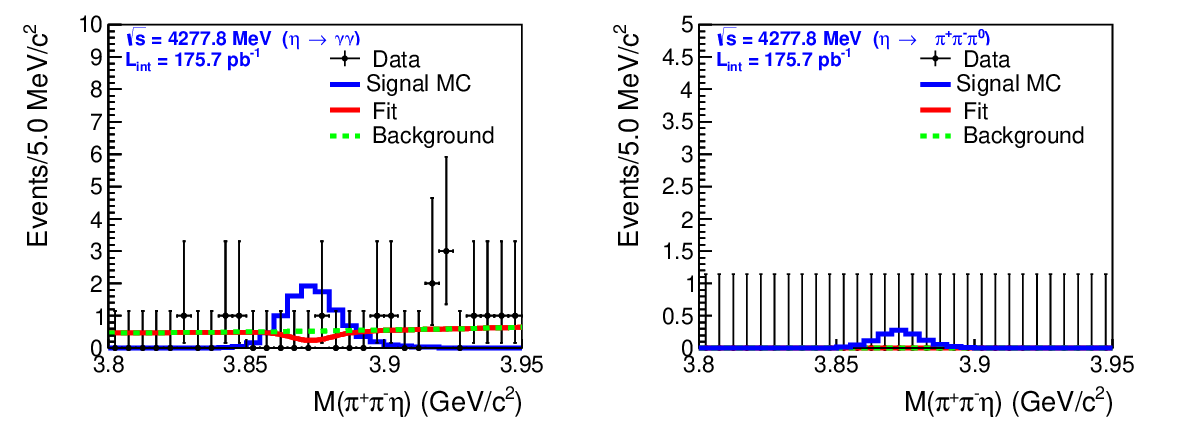}
	\includegraphics[width=0.48\textwidth, height=0.21\textwidth]{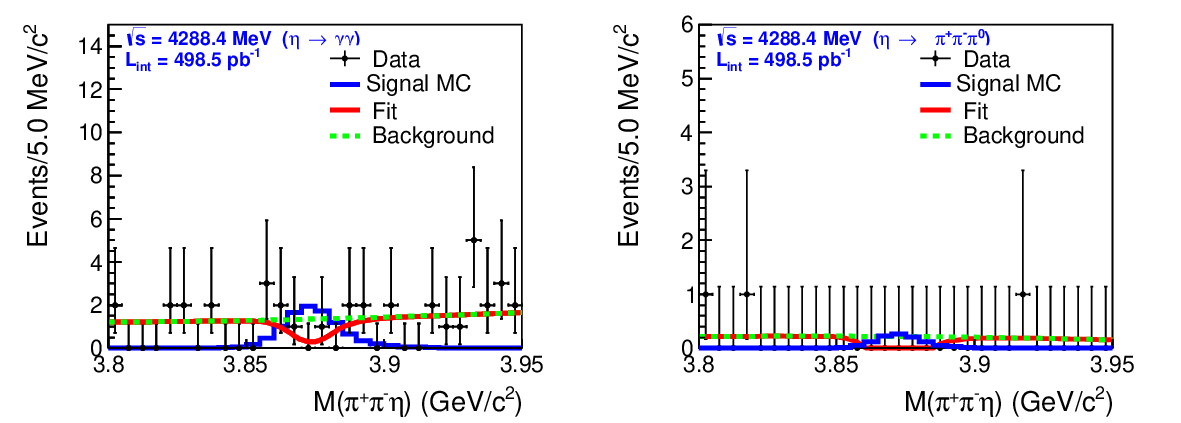}
	\includegraphics[width=0.48\textwidth, height=0.21\textwidth]{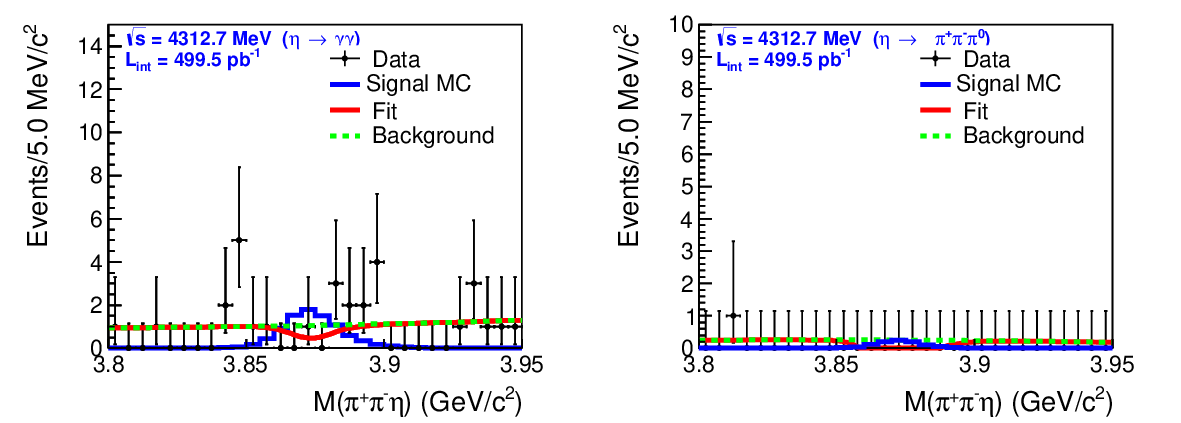}
	\includegraphics[width=0.48\textwidth, height=0.21\textwidth]{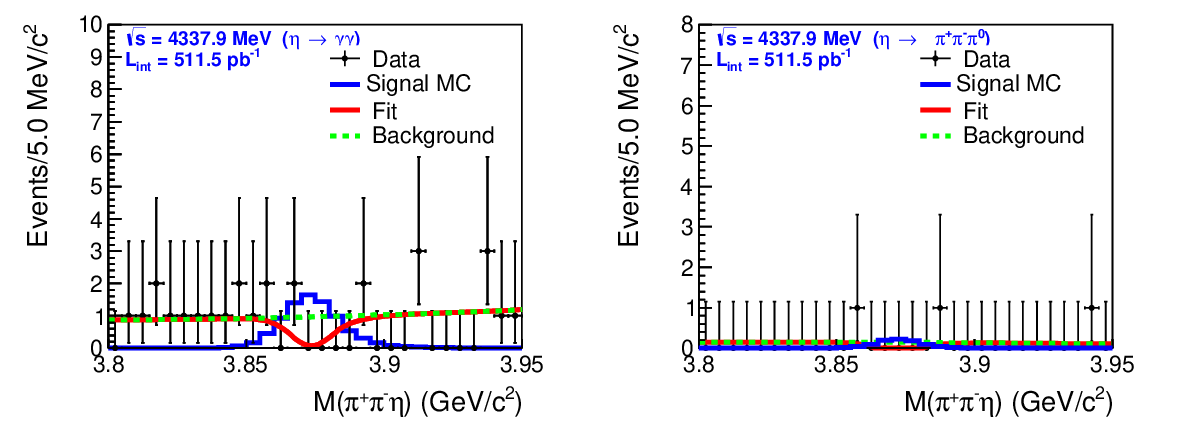}
	\caption{The fit results at c.m. energies ranging from $\sqrt{s}=$ 4258.0 to 4337.9 MeV. The signal MC shape convolved with a Gaussian function is used to fit the invariant mass spectrum of $\pi^{+}\pi^{-}\eta$ at each energy. Black dots with error bars are data, the red solid curve shows the fit result, the blue dashed curve is the signal shape with arbitrary normalization and the green dashed curve is the background contribution.}
	\label{fit fig2}
\end{figure*}

\section{Systematic uncertainty at each c.m. energy}
\label{sys_ecm}
\begin{table*}[!htbp]
\caption{Systematic uncertainties on the product of  $(\sigma[e^{+}e^{-}\rightarrow\gamma \chi_{c1}(3872)]  \mathcal{B}[\chi_{c1}(3872)\rightarrow\pi^{+}\pi^{-}\eta])^{\rm up}$. $\Delta$ denotes the total systematic uncertainty.  All entries are in percentage. }
	\hspace{15pt}
	\centering
	\footnotesize
	\setlength{\tabcolsep}{2pt}
	\begin{tabular}{c c c c c c c c c c }
		\hline
		\hline
		$\sqrt{s}$~(MeV) & Photon & Tracking & Luminosity & Branching fraction & Kinematic fit &  Decay model & ISR & $\eta(\pi^{0})$ reconstruction &$\Delta$\\
		\hline
		4128.8$\pm$0.4&3.0&2.5&0.7&0.7&2.0&5.6&0.3&1.0&7.2\\
		4157.8$\pm$0.3&3.0&2.5&0.7&0.7&1.8&5.6&0.6&1.0&7.2\\
		4178.0$\pm$0.8&3.0&2.5&0.7&0.7&2.1&6.0&0.9&1.0&7.6\\
		4189.1$\pm$0.3&3.0&2.5&0.7&0.7&1.9&5.2&1.1&1.0&7.0\\
		4199.2$\pm$0.3&3.0&2.5&0.7&0.7&2.0&6.2&1.3&1.0&7.8\\
		4209.4$\pm$0.3&3.0&2.5&0.7&0.7&1.9&5.0&1.2&1.0&6.9\\
		4218.9$\pm$0.3&3.0&2.5&0.7&0.7&1.8&6.4&1.0&1.0&7.9\\
		4226.3$\pm$0.7&3.0&2.5&0.7&0.7&2.0&5.8&1.0&1.0&7.5\\
		4235.8$\pm$0.3&3.0&2.5&0.7&0.7&2.0&6.2&1.1&1.0&7.8\\
		4244.0$\pm$0.3&3.0&2.5&0.7&0.7&1.8&6.0&0.9&1.0&7.6\\
		4258.0$\pm$0.7&3.0&2.5&0.7&0.7&1.7&5.2&0.9&1.0&6.9\\
		4266.8$\pm$0.3&3.0&2.5&0.7&0.7&2.1&4.8&0.9&1.0&6.7\\
		4277.8$\pm$0.5&3.0&2.5&0.7&0.7&2.0&4.1&0.8&1.0&6.2\\
		4288.4$\pm$0.3&3.0&2.5&0.7&0.7&2.0&6.2&0.7&1.0&7.7\\
		4312.7$\pm$0.4&3.0&2.5&0.7&0.7&1.7&5.8&0.6&1.0&7.4\\
		4337.9$\pm$0.4&3.0&2.5&0.7&0.7&1.7&5.6&0.4&1.0&7.2\\
		\hline
		\hline
	\end{tabular}
	\label{Sum of sys err}
\end{table*}
\end{widetext}

\end{document}